\documentclass[12pt]{article}

\usepackage{amsmath}

\usepackage{amsfonts}

\usepackage{amssymb}

\usepackage[latin1]{inputenc}

\usepackage{graphics}

\usepackage{graphicx}

\usepackage{epsfig}

\usepackage{amsthm}

\usepackage{lineno}

\usepackage{mathrsfs}

\usepackage{latexsym}

\usepackage{float}

\usepackage{mathtools}

\usepackage[T1]{fontenc}

\usepackage{ae}

\usepackage{authblk}

\usepackage{latexsym}

\usepackage[colorlinks,citecolor=blue,urlcolor=blue,linkcolor=blue]{hyperref}

\usepackage{xcolor}

\newenvironment{destaque}{\begin{quotation}\small}{\end{quotation}}
\usepackage[figurename=Fig.]{caption}
\usepackage{subcaption}
\usepackage{titlesec}
\titlelabel{\thetitle.\quad}

\newcommand{\abs}[1] {\lvert #1 \rvert}
\newcommand{\mvl}[1] {\left\langle #1 \right\rangle}
\newcommand{\ket}[1] {\lvert #1 \rangle}

\newcommand{\bkt}[2] {\left\langle #1 \mid #2 \right\rangle}
\newcommand{\xvl}[2] {\left\langle #1 \rvert #2 \lvert #1 \right\rangle}

\date{}

\begin{document}

\title{\bf Generalized Uncertainty Principle for Entangled States of Two Identical Particles}

\author[1]{K.~C.~Lemos Filho\thanks{kim.vasco@gmail.com}}
\author[2]{B. B. Dilem\thanks{bernardob@ifes.edu.br}}
\author[1]{R. O. Francisco\thanks{ronald.francisco@ufes.br}}
\author[1,3]{J. C. Fabris\thanks{julio.fabris@cosmo-ufes.org}}
\author[1]{J. A. Nogueira\thanks{jose.nogueira@ufes.br}}
\affil[1]{Universidade Federal do Esp\'{\i}rito Santo -- Ufes\\ Vit\'oria, Esp\'{\i}rito Santo, 29.075-910, Brasil}
\affil[2]{Instituto Federal do Esp\'{\i}rito Santo -- Ifes\\ Alegre, Esp\'{\i}rito Santo, 29.520-000, Brasil}
\affil[3]{National Research Nuclear University MEPhI, Kashirskoe sh. 31, Moscow 115409, Russia}

\maketitle

\begin{abstract}
\begin{destaque}

In this work we determine the consequences of the quantum entanglement of a system of two identical particles when a generalized uncertainty principle (GUP) is considered. GUP's are usually associated with the existence of a minimal length. We focus on the main GUP's (KMM, ADV, Pedram and Nouicer) and then we determine the minimal uncertainties in position induced by those modified GUP's. Our results point out that the minimal uncertainty is reduced by half of its usual value independently of the GUP employed. This implies that the minimal length is also reduced by half. On the other hand, it is generally expected that the minimal length must not depend on physical system. We overcome this apparent paradox by realizing that the entangled system is composed by two particles so that an effective parameter related to the minimal length must be employed.
\end{destaque}
\end{abstract}

\pagenumbering{arabic}


\section{Introduction}
\label{introd}

Amongst the (many) new concepts introduced by the quantum mechanics, the quantum entanglement \cite{Horodecki:2009,Brody} is one that, probably, more contradicts our common sense. Although in the beginning the quantum entanglement were only associated to theoretical aspects of the quantum mechanics, specially those related to the non-locality or the complementarity (hidden variables) \cite{EPR:1935}, nowadays it is a key component of the applications and experiments on quantum information, quantum computation and quantum teleportation \cite{Nielsen:2002,Nielsen:2010}.

The uncertainty principle is one of the fundamental cornerstones of the quantum mechanics. Nevertheless, it is a principle: its correct form can not be proven. This opens the way to the possibility that its canonical form, described by the Heisenberg's uncertainty principle (HUP), can be generalized\footnote{For more information on generalization of the Heisenberg's uncertainty principle see Refs.~\cite{konishi:1990,Maggiore1:1993,Maggiore2:1993,Maggiore:1994,Adler:1999,Scardigli:1999,Scardigli:2003,Vagenas:2009,Tawfik:2014,Scardigli:2015,Tawfik2:2015}.}. Possible generalizations of the HUP, whose origin can be traced back to quantum gravity, are given by introducing a non-zero minimal uncertainty in the measurement of position. The non-zero minimal uncertainty in position is, therefore, understood as a minimal-length scale, below which the necessary amount of energy to probe the position of a particle is so hight that it disturbs the space-time so that the concept of a length measurement loses its meaning. Hence, theories searching to describe a quantum approach for gravity lead generally to the existence of a minimal length. In fact, a minimal length actually appears in almost all proposed theories of quantum gravity. For this reason theories formulated in a minimal-length scenario are considered to be effective theories of quantum gravity \cite{Hossenfelder:2006,Kober1:2012,Nouicer1:2012,Ong2:2018}.

In 1999, Yoon-Ho Kim and Yanhua Shih conducted an experiment whose results apparently suggested a violation of the HUP \cite{Kim:1999}. However, G. Rigolin pointed out that there is, in fact, no violation of the HUP, because the HUP is derived for particles that are non-correlated (non-entangled). In the Kim and Shih's experiment, the photons of the pair are correlated (entangled) when one of the physical slits is replaced by a virtual slit\footnote{The interaction of the photon with a physical slit destroys the correlation between the photons of the pair.}. This indicates that the canonical HUP is no longer directly applicable due to the modification caused by quantum entanglement.\cite{Rigolin:2002,Rigolin:2016}

An immediate question arising from the previous considerations is how the quantum entanglement modifies a generalized uncertainty principle (GUP), in others words, what is the effect caused by quantum entanglement at the minimal-length scale. The answer for that question is important in order to know the role of the quantum entanglement at the minimal-length scale (maybe appearing at the Planck scale) or in the early Universe. Unfortunately, this issue has been little considered in the literature.  G. Blado, F. Herrera and J. Erwin \cite{Blado:2018} have studied the inseparability conditions with the most usual GUP correction, whereas  D. Park\cite{Park:2022} has used a coupled harmonic oscillator in order to find the effects of the quantum entanglement with a linear GUP.

The purpose of this work is to answer that question considering the main proposals of generalization for the HUP which take into account the existence of a minimal length in nature. With this goal in mind, we will analyze the modifications in the HUP arising from the quantum entanglement of two identical particles determining the minimal uncertainty associated to them.

The outline of this paper is as follows. In Section \ref{UPES} we obtain an expression of the uncertainty principle for entangled states which is independent of the chosen (generalized) uncertainty principle. In Section \ref{UPESDMLS} we find the modified uncertainty principle for a pair of identical particles regarding the main proposals of GUP's: Kempf, Mangano and Mann GUP (KMM-GUP), Ali, Das and Vagenas GUP (ADV-GUP), Pedram GUP and Nouicer GUP. In Section \ref{UBMLV} we estimate an upper bound for the minimal-length value. We present our conclusions in the Section \ref{Concl}.


\section{Uncertainty principle for entangled states}
\label{UPES}

The Hilbert space of the state vectors ${\mathcal E}$ of a system of $N$ particles is given by the tensor product of the Hilbert spaces of the state vectors ${\mathcal E}_{i}$ of each particle \cite{Cohen,Messiah},
\begin{equation}
	{\cal E} = {\cal E}_{1} \otimes \dots \otimes {\cal E}_{N}.
\end{equation}
The position and momentum linear operators of the i-th particle, which act on the state vectors $\ket{\psi} \in \cal{E}$, are the extensions $\tilde{Q}_{i}$ and $\tilde{P}_{i}$ defined as
\begin{equation}
	\tilde {Q}_{i} = \Bbb{I}_{1} \otimes \dots \otimes \hat{x}_{i} \otimes \dots \otimes \Bbb{I}_{N},
\end{equation}
\begin{equation}
	\tilde {P}_{i} = \Bbb{I}_{1} \otimes \dots \otimes \hat{p}_{i} \otimes \dots \otimes \Bbb{I}_{N},
\end{equation}
where $\Bbb{I}_{i}$ is the identity operator in ${\cal E}_{i}$ and $\hat{x}_{i}$ and $\hat{p}_{i}$ are the position and the momentum operators of the i-th particle acting on the state vectors $\ket{\psi_{i}} \in {\cal E}_{i}$.

The extensions $\tilde{Q}_{i}$ and $\tilde{P}_{i}$ do not satisfy the canonical uncertainty principle (HUP), because $\tilde{Q}_{i}$ and $\tilde{P}_{i}$ are not physical observables \cite{Rigolin:2002,Rigolin:2016,Cohen,Messiah}. Physical observables are operators which commute with every permutation operators of the particles system. Hence, the operators $\tilde{Q}$ and $\tilde{P}$, defined as
\begin{equation}
	\tilde{Q} := \sum_{i=1}^{N} \tilde{Q}_{i}
\end{equation}
and
\begin{equation}
	\tilde{P} := \sum_{i=1}^{N} \tilde{P}_{i},
\end{equation}
are physical observables and they satisfy the relation
\begin{equation}
	\label{cr}
	\left(\Delta Q \right)^{2} \left( \Delta P \right)^{2} \geq \frac{1}{4}  \left| \langle [\tilde{Q},\tilde{P}]  \rangle \right|^{2}.
\end{equation}
The relation (\ref{cr}) is general. It does not depend on whether the system of particle is entangled or not.

As it was showed by G. Rigolin \cite{Rigolin:2002,Rigolin:2016}, if the state of the particles system is entangled then the operators $\tilde{Q}_{i}$ and $\tilde{P}_{i}$ do not satisfy the canonical Heisenberg uncertainty principle (HUP) - as previously stated.

Next, we briefly review the Rigolin's result for a two particles system.\\
Initially, we define the functions
\begin{eqnarray}
	\label{CQ}
	C_{Q}(1,2) &:=& \mvl{\tilde{Q}_{1} \tilde{Q}_{2}} - \mvl{\tilde{Q}_{1}}\mvl{\tilde{Q}_{2}},\\
	\label{CP}
	C_{P}(1,2) &:=& \mvl{\tilde{P}_{1} \tilde{P}_{2}} - \mvl{\tilde{P}_{1}}\mvl{\tilde{P}_{2}},
\end{eqnarray}
which are called {\it quantum covariance functions} (QCF). By definition, QCF's vanish if and only if the system is separable \cite{Torre:1989}. Therefore (\ref{CQ}) and (\ref{CP}) are zero for any not entangled quantum system. From the definitions of $\Delta Q$ and $\Delta P$ we have\footnote{Note that $[\tilde{Q}_{1},\tilde{Q}_{2}] = 0$ and $[\tilde{P}_{1},\tilde{P}_{2}] = 0$.}
\begin{equation}
	\left( \Delta_{\psi} Q \right)^{2} = \xvl{\psi}{\tilde{Q}^{2}} - \xvl{\psi}{\tilde{Q}}^{2}.
\end{equation}
From now on, we omit the subscript $\psi$ for the sake of simplicity, whenever this  does not cause any confusion. Thus,
\begin{equation}
	\label{DQ}
	\left( \Delta Q \right)^{2} = \left( \Delta Q_{1} \right)^{2} + \left( \Delta Q		    _{2} \right)^{2} + 2 \left(\mvl{\tilde{Q}_{1} \tilde{Q}_{2}} - \mvl{\tilde{Q}_{1}}\mvl{\tilde{Q}_{2}} \right).
\end{equation}
In the same way,
\begin{equation}
	\label{DP}
	\left( \Delta P \right)^{2} = \left( \Delta P_{1} \right)^{2} + \left( \Delta P_{2} \right)^{2} + 2 \left(\mvl{\tilde{P}_{1} \tilde{P}_{2}} - \mvl{\tilde{P}_{1}}\mvl{\tilde{P}_{2}} \right).
\end{equation}
Using the results (\ref{DQ}) and (\ref{DP}) into Eq. (\ref{cr}) we obtain
\begin{equation*}
	\left[ \left( \Delta Q_{1} \right)^{2} + \left( \Delta Q_{2} \right)^{2} + 2 	  \left(\mvl{\tilde{Q}_{1} \tilde{Q}_{2}} - \mvl{\tilde{Q}_{1}}\mvl{\tilde{Q}_{2}} \right) \right] \times
\end{equation*}
\begin{equation}
	\label{UP1}
	\left[ \left( \Delta P_{1} \right)^{2} + \left( \Delta P_{2} \right)^{2} + 2 \left(\mvl{\tilde{P}_{1} \tilde{P}_{2}} - \mvl{\tilde{P}_{1}}\mvl{\tilde{P}_{2}} \right) \right] \geq \frac{1}{4}  \left| \langle [\tilde{Q},\tilde{P}]  \rangle \right|^{2}.
\end{equation}

Using the QCF's, Eqs. (\ref{CQ}) and (\ref{CP}), we have
\begin{equation}
	\label{UP2}
	\left[ \left( \Delta Q_{1} \right)^{2} + \left( \Delta Q_{2} \right)^{2} + 2C_{Q}(1,2) \right] \left[ \left( \Delta P_{1} \right)^{2} + \left( \Delta P_{2} \right)^{2} + 2C_{P}(1,2) \right] \geq \frac{1}{4}  \left| \langle [\tilde{Q},\tilde{P}]  \rangle \right|^{2},
\end{equation}
or
\begin{equation}
	\label{UP3}
	\sum_{i,j=1}^{2} C_{Q}(i,j) \sum_{k,l=1}^{2} C_{P}(k,l) \geq \frac{1}{4}  \left| \langle [\tilde{Q},\tilde{P}]  \rangle \right|^{2},
\end{equation}
since $C_{Q}(i,i) = \left( \Delta Q_{i} \right)^{2}$, $C_{P}(i,i) = \left( \Delta P_{i} \right)^{2}$, $C_{Q}(i,j) = C_{Q}(j,i)$ and $C_{P}(i,j) = C_{P}(j,i)$.

In this work, we concern with the case of an entangled system of two identical particles, so we are going to handle Eq. (\ref{UP2}) in order to express it in a more  appropriate way. For this end, we define
\begin{equation}
	\ket{\psi\prime} := \left( \tilde{Q}_{1} - \tilde{Q}_{2} \right) \ket{\psi},
\end{equation}
with $\ket{\psi\prime}, \ket{\psi} \in \cal{E}$ and $\bkt{\psi}{\psi} = 1$. 
Therefore,
\begin{equation}
	\bkt{\psi\prime}{\psi\prime} = \left( \Delta_{\psi} Q_{1} \right)^{2} + \left( \Delta_{\psi} Q_{2} \right)^{2} - 2 \mvl{\tilde{Q}_{1} \tilde{Q}_{2}}_{\psi} + \mvl{\tilde{Q}_{1}}^{2}_{\psi} + \mvl{\tilde{Q}_{2}}^{2}_{\psi}.
\end{equation}
Now, using the Schwarz inequality, $\bkt{\psi}{\psi} \bkt{\psi\prime}{\psi\prime} \geq \bkt{\psi}{\psi\prime}\bkt{\psi\prime}{\psi}$, we have
\begin{equation}
	\label{iq1}
	\left( \Delta_{\psi} Q_{1} \right)^{2} + \left( \Delta_{\psi} Q_{2} \right)^{2} \geq 2\left(\mvl{\tilde{Q}_{1} \tilde{Q}_{2}}_{\psi} - \mvl{\tilde{Q}_{1}}_{\psi} \mvl{\tilde{Q}_{2}}_{\psi} \right).  
\end{equation}
In the same way,
\begin{equation}
	\label{iq2}
	\left( \Delta_{\psi} P_{1} \right)^{2} + \left( \Delta_{\psi} P_{2} \right)^{2} \geq 2\left(\mvl{\tilde{P}_{1} \tilde{P}_{2}}_{\psi} - \mvl{\tilde{P}_{1}}_{\psi} \mvl{\tilde{P}_{2}}_{\psi} \right).  
\end{equation}
Finally, from inequalities (\ref{iq1}), (\ref{iq2}) and (\ref{UP2}) we obtain
\begin{equation}
	\label{UP4}
	\left[ \left( \Delta Q_{1} \right)^{2} + \left( \Delta Q_{2} \right)^{2} \right] \left[ \left( \Delta P_{1} \right)^{2} + \left( \Delta P_{2} \right)^{2} \right] \geq \frac{1}{16}  \left| \langle [\tilde{Q},\tilde{P}]  \rangle \right|^{2}.
\end{equation}

In the case where $\left( \Delta Q_{1} \right)^{2} = \left( \Delta Q_{2} \right)^{2}$ and $\left( \Delta P_{1} \right)^{2} = \left( \Delta P_{2} \right)^{2}$ the inequality (\ref{UP4}) becomes
\begin{equation}
	\label{UP5}
	\Delta Q_{i} \Delta P_{i} \geq \frac{1}{8}  \left| \langle [\tilde{Q},\tilde{P}]  \rangle \right|.
\end{equation}

It is worth noting that the expression of the inequality (\ref{UP5}) is independent of the chosen uncertainty principle that does not take into account the quantum correlation. This uncertainty principle is related to the commutation relation $ [\tilde{Q},\tilde{P}] $.


\section{Uncertainty principle for entangled states in different minimal-length scenarios}
\label{UPESDMLS}

In this section we consider a system of two entangled identical particles whose momenta have the same value but opposite directions, that is, $\vec{p}_{1} = - \vec{p}_{2}$, just as in the Kim and Shih's experiment \cite{Kim:1999}. Therefore, in this case\footnote{Note that $p_{1} = \abs{\vec{p}_{1}}$ and $p_{2} = \abs{\vec{p}_{2}}$, therefore $p_{1} = p_{2}$.} $\mvl{\hat{p}_{1}} = \mvl{\hat{p}_{2}}$. Moreover, such a consideration also allows us to estimate, in the next section, an upper bound for the value of the minimal length based on the experimental results obtained by Kim and Shih.

\subsection{Heisenberg uncertainty principle}

Before we consider a minimal-length scenario it is appropriate to determine the change in the canonical HUP, that is, in a scenario in which effects of quantum gravity are not present. The canonical HUP for states of one simple-particle is
\begin{equation}
\label{HUP}
\Delta x \Delta p \geq \frac{\hbar}{2}.
\end{equation}
The commutation relation related to the HUP is
\begin{equation}
\label{CRHUP}
[\hat{x}, \hat{p}] =  i \hbar.
\end{equation}
Hence\footnote{Note that $[\tilde{Q}_{i}, \tilde{P}_{j}] = 0$, for $i \neq j$.}
\begin{equation}
\label{CR1}
[\tilde{Q}, \tilde{P}] = [\tilde{Q}_{1} + \tilde{Q}_{2}, \tilde{P}_{1} + \tilde{P}_{2}] = 2 i \hbar.
\end{equation}
Substituting Eq. (\ref{CR1}) into Eq. (\ref{UP5}), and assuming\footnote{We are going to assume the same for the next GUP's.} $\left(\Delta Q_{1} \right)^{2} = \left(\Delta Q_{2} \right)^{2}$ and $\left(\Delta P_{1} \right)^{2} = \left(\Delta P_{2} \right)^{2}$,  we obtain
\begin{equation}
	\label{EPHUP}
	\Delta Q_{i} \Delta P_{i} \geq \frac{\hbar}{4},
\end{equation}
where $i = 1, 2$.
The result (\ref{EPHUP}) shows that for a system of two entangled identical particles the HUP is modified. Such an outcome is not new, it was already obtained by G. Rigolin in 2002 \cite{Rigolin:2002} and then in 2016 \cite{Rigolin:2016}.

From a quick glance at the result (\ref{EPHUP}) and recalling that dimensionally $\left( \Delta Q \right)_{min} \propto \hbar$, we expect the minimal uncertainty in the position will be reduced by half for all GUP's.

\subsection{KMM GUP}

The GUP
\begin{equation}
	\label{KMMGUP}
	\Delta x_{i} \Delta p_{i} \geq \frac{\hbar}{2} \left[ 1 + \beta \left( \Delta p_{i} \right)^{2} + \beta \mvl{\hat{p}_{i}}^{2} \right],
\end{equation}
where $\beta$ is a parameter related to the minimal length, has been  proposed by A. Kempf, G. Mangano an R. B. Mann (KMM-GUP) \cite{Kempf1:1995} and it is the most used in the literature. The commutation relation related to it is given by
\begin{equation}
\label{KMMCR}
[\hat{x}_{i}, \hat{p}_{i}] =  i \hbar \left( 1 + \beta \hat{p}_{i}^{2} \right).
\end{equation}
Hence
\begin{equation}
\label{CR2}
[\tilde{Q}, \tilde{P}] = i \hbar \left[2 + \beta \left( \tilde{P}_{1}^{2} + \tilde{P}_{2}^{2} \right) \right] = 2i \hbar \left( 1 + \beta \tilde{P}_{i}^{2} \right).
\end{equation}
Substituting Eq. (\ref{CR2}) into Eq. (\ref{UP5}) we get
\begin{equation}
	\label{EKMMGUP}
	\Delta Q_{i} \Delta P_{i} \geq \frac{\hbar}{4} \left[ 1 + \beta \left( \Delta P_{i} \right)^{2} + \gamma \right],
\end{equation}
where $\gamma := \beta \mvl{\tilde{P}_{i}}^{2}$.

The modified KMM-GUP (\ref{EKMMGUP}) induces the existence of a minimal uncertainty given by
\begin{equation}
	\label{MLKMM}
	\left( \Delta Q_{i} \right)_{min} = \frac{\hbar}{2} \sqrt{\beta}.
\end{equation}

The result above shows that the non-zero minimal uncertainty in position induced by the KMM-GUP for two entangled identical particles is twice smaller than for a separable system of two identical particles (non-entangled).

\subsection{ADV GUP}

A. Farag Ali, S. Das and E. C. Vagenas have proposed a GUP related to a commutation relation which has a linear and a quadratic term in the momentum operator \cite{Farag:2009},
\begin{equation}
\label{ADVCR}
[\hat{x}_{i}, \hat{p}_{i}] =  i \hbar \left( 1 - 2\alpha\hat{p}_{i} + 4\alpha^{2}\hat{p}_{i}^{2} \right),
\end{equation}
where $\alpha$ is a parameter related to the minimal length. Besides the existence of a minimal length this linear approach induces a maximal uncertainty in the momentum, too.
Then, from Eq. (\ref{ADVCR}) we get
\begin{equation}
\label{CR3}
[\tilde{Q}, \tilde{P}] = i \hbar \left[2 - 2\alpha \left( \tilde{P}_{1} + \tilde{P}_{2} \right) + 4\alpha^{2}\left( \tilde{P}_{1}^{2} + \tilde{P}_{2}^{2} \right) \right].
\end{equation}
Therefore,
\begin{equation}
	\langle [\tilde{Q},\tilde{P}]  \rangle = i \hbar \left[2 - 2\alpha \left( \mvl{\tilde{P}_{1}} + \mvl{\tilde{P}_{2}} \right) + 4\alpha^{2}\left( \mvl{\tilde{P}_{1}^{2}} + \mvl{\tilde{P}_{2}^{2}} \right) \right],
\end{equation}
\begin{equation}
	\label{CR4}
	\langle [\tilde{Q},\tilde{P}]  \rangle = 2 i \hbar \left[1 - 2\alpha\mvl{\tilde{P}_{i}}   + 4\alpha^{2}\mvl{\tilde{P}_{i}^{2}} \right].
\end{equation}
Substituting Eq. (\ref{CR4}) into Eq. (\ref{UP5}) we obtain
\begin{equation}
	\label{EADVGUP}
	\Delta Q_{i} \Delta P_{i} \geq \frac{\hbar}{4} \left[ 1 + 4\alpha^{2} \left( \Delta P_{i} \right)^{2} + \gamma\prime \right],
\end{equation}
where $\gamma\prime := - 2\alpha\mvl{\tilde{P}_{i}} + 4\alpha^{2} \mvl{\tilde{P}_{i}}^{2}$.

The modified ADV-GUP (\ref{EADVGUP}) induces the existence of a non-zero minimal uncertainty given by
\begin{equation}
	\label{MLKMM}
	\left( \Delta Q_{i} \right)_{min} = \hbar \alpha,
\end{equation}
which once again is twice smaller than for a non-entangled system of two particles.

It is important to note that the maximal uncertainty in the momentum is not changed, $\left( \Delta P_{i} \right)_{max} = \frac{1}{2 \alpha} $. But this was to be expected because $\left( \Delta P_{i} \right)_{max}$ does not depend on $\hbar$.

\subsection{Pedram GUP}

With the intention of remedying some problems arising from KMM-GUP and ADV-GUP - such as the absence of a maximal momentum required in theories of doubly special relativity (DSR) and of commutative geometry, and also the lack of a valid approach for all order in the parameter related to the minimal length - P. Pedram has proposed a GUP \cite{Pedram1:2012,Pedram2:2012} based on the commutation relation given  by
\begin{equation}
\label{PCR}
[\hat{x}_{i}, \hat{p}_{i}] =  \frac{i \hbar}{1 - \beta \hat{p}_{i}^{2}}.
\end{equation}
Hence
\begin{equation}
\label{CR5}
[\tilde{Q}, \tilde{P}] = \frac{i \hbar}{1 - \beta \tilde{P}_{1}^{2}} + \frac{i \hbar}{1 - \beta \tilde{P}_{2}^{2}}.
\end{equation}
Using the so-called Jensen's Inequality \cite{Jensen:1906} we have
\begin{equation}
	\label{CR6}
	\langle [\tilde{Q},\tilde{P}]  \rangle \geq \frac{i \hbar}{1 - \beta \mvl{\tilde{P}_{1}^{2}}} +\frac{i \hbar}{1 - \beta \mvl{\tilde{P}_{2}^{2}}}^.
\end{equation}
Substituting Eq. (\ref{CR6}) into Eq. (\ref{UP5}) we obtain
\begin{equation}
	\label{EPGUP}
	\Delta Q_{i} \Delta P_{i} \geq \frac{\hbar}{4}\frac{1}{ \left[1 - \beta \left( \Delta P_{i} \right)^{2} - \gamma \right]},
\end{equation}
where again $\gamma := \beta \mvl{\tilde{P}_{i}}^{2}$.

Therefore, the modified Pedram-GUP (\ref{EPGUP}) introduces a non-zero minimal uncertainty given by
\begin{equation}
	\label{MLKMM}
	\left( \Delta Q_{i} \right)_{min} = \frac{3 \hbar}{8} \sqrt{3 \beta},
\end{equation}
which is also twice smaller than for a non-entangled system of a pair of identical particles.

\subsection{Nouicer GUP}

In the canonical field theory, in the context of non-commutative coherent states representation and field theory on non-anticommutative superspace the Feynman propagator displays an ultra-violet (UV) cut-off of the form $e^{-\beta p^{2}}$ \cite{Moffat:2001,Smailagic1:2003,Smailagic2:2003,Smailagic1:2004}. In consequence, K. Nouicer has proposed an exponential all orders GUP \cite{Nouicer1:2007,Nouicer2:2007} based on the commutation relation given by 
\begin{equation}
\label{expCR}
[\hat{x}_{i}, \hat{p}_{i}] =  i \hbar e^{\beta \hat{p}_{i}^{2}}.
\end{equation}
Hence
\begin{equation}
\label{CR7}
[\tilde{Q}, \tilde{P}] = i \hbar \left( e^{\beta \tilde{P}_{1}^{2}} + e^{\beta \tilde{P}_{2}^{2}} \right).
\end{equation}
Again, using the so-called Jensen's Inequality \cite{Jensen:1906} we have
\begin{equation}
	\label{CR8}
	\langle [\tilde{Q},\tilde{P}]  \rangle \geq i \hbar \left( e^{\beta \mvl{\tilde{P}_{1}^{2}}} + e^{\beta \mvl{\tilde{P}_{2}^{2}}} \right).
\end{equation}
Substituting Eq. (\ref{CR8}) into Eq. (\ref{UP5}) we get
\begin{equation}
	\label{EexpGUP}
	\Delta Q_{i} \Delta P_{i} \geq \frac{\hbar}{4} e^{ \beta \left( \Delta P_{i} \right)^{2} + \gamma},
\end{equation}
where once again $\gamma := \beta \mvl{\tilde{P}_{i}}^{2}$.

Therefore, the modified Nouicer-GUP (\ref{EexpGUP}) introduces a minimal uncertainty given by
\begin{equation}
	\label{MLKMM}
	\left( \Delta Q_{i} \right)_{min} = \frac{\hbar}{2} \sqrt{\frac{e \beta}{2}}.
\end{equation}
Finally, we note that $\left( \Delta Q_{i} \right)_{min}$ is also twice smaller than for a non-entangled system of a pair of identical particles.

\subsection{Minimal length}

If a non-zero minimal uncertainty in position can be interpreted as a minimal length then the previous results show that the minimal length for two entangled identical particles can be twice smaller than for a separable system. It is clear this statement must be taken with care because a minimal length should be a constant, that is, it must not depend on the physical system, it is a quantum gravitation effect. In fact, the minimal length should be an invariant as well as the light speed is. The answer for that apparent contradiction is possibly because the system is made up of two particles. C. Quesne and V. M. Tkachuck\cite{Quesne:2010} have claimed that for a system composed by $N$ particles the effective parameter related to the minimal length, $\beta$, is reduced by a factor $\frac{1}{N^{2}}$. Hence, $\frac{\hbar}{2} \sqrt{\beta}$ is not the correct minimal uncertainty in position $\left( \Delta Q_{i} \right)_{min}$ for the modified KMM-GUP, but $\frac{\hbar}{2} \sqrt{\beta_{i}}$, where \cite{Quesne:2010,Nogueira:2016}
\begin{equation}
\beta = \frac{\beta_{i}}{2^{2}}.
\end{equation}
Consequently, we find that $\left( \Delta Q_{i} \right)_{min} = \hbar \sqrt{\beta}$, therefore $l_{min} = \hbar \sqrt{\beta}$, as we expected\footnote{It is worth noting that if the minimal uncertainty was greater for the entangled system, we could suppose that the quantum entanglement decreases the accuracy of a position measurement thereby increasing the minimal uncertainty.}. In the same way for others GUP's.

The reader might want to claim that the minimal length is actually described by the minimal uncertainty of the entangled system. However, according to G. Rigolin\cite{Rigolin:2016} we can presume that the minimal uncertainty for a system of $N$ entangled identical particles is reduced by $\frac{1}{N}$. Since, in principle, there is no limit for the number of particles for a entangled system, there would also be no limit for the minimal length.


\section{Upper bound for the minimal-length value}
\label{UBMLV}

In the Kim and Shih's experiment entangled identical pairs of photons were produced by spontaneous parametric down conversion (SPDC) with momentum conservation. A narrow physical slit was placed along the trajectory of one of the photons, whereas the other photon of the pair (called 2) passed through a virtual slit. The ghost image experimental technique \cite{Pittman:1995} ensured that the quantum correlation between the pair of photons was not destroyed. Then simultaneous detection of photons of the pairs were performed and data just for coincidence events were obtained in case when the photon 2 passed through a virtual slit (non-slit case) and in case when the photon 2 passed through a physical slit (slit case).

From the experimental data obtained by Kim and Shih one gets that \cite{Rigolin:2016}
\begin{equation}
	\label{rate}
	\frac{\Delta P_{2}^{ns}}{\Delta P_{2}^{s}} = \frac{1.25}{2.15},
\end{equation}
where $\Delta P_{2}^{ns}$  is uncertainty in momentum of the photon 2 in the non-slit case and $\Delta P_{2}^{s}$  is uncertainty in momentum of the photon 2 in slit case. Since the width of the slit was 0.16 mm ($\Delta Q_{2} = 0.16$ mm), the uncertainty in momentum in the slit case for KMM-GUP can be find from\footnote{Note that in with-slit case the correlation between the photons of the pair was destroyed, therefore the photons were not entangled and the usual KMM-GUP, Eq. (\ref{KMMGUP}), is held.}
\begin{equation}
	\label{eq1}
	0.16  \Delta P^{s}_{2} = \frac{\hbar}{2} \left[ 1 + \beta \left( \Delta P^{s}_{i} \right)^{2} + \gamma \right].
\end{equation}
Eq. (\ref{eq1}) has real roots only if $\beta \leq \frac{0.16^{2}}{\hbar^{2} \eta}$, where $\eta := 1 + \gamma$. Thus,
\begin{equation}
	\label{r1}
	\Delta P_{2+}^{s} = \frac{0.32}{\hbar \beta} - \frac{\hbar \eta}{0.32} - \beta \frac{\hbar^{3}\eta^{2}}{0.32^{3}}
\end{equation}
and
\begin{equation}
	\label{r2}
	\Delta P_{2-}^{s} = \frac{\hbar \eta}{0.32}  +  \beta \frac{\hbar^{3} \eta^{2}}{0.32^{3}}.
\end{equation}

Now, using the above results we can estimate an upper bound for the minimal-length value induced by KMM-GUP. Hence, substituting the root (\ref{r1}) into Eq. (\ref{EKMMGUP}) we have that
\begin{equation}
\beta \leq \frac{3.58 \times 10^{-2}}{\hbar^{2} \eta}.
\end{equation}
Therefore,
\begin{equation}
l_{min} \leq \hbar \sqrt{\frac{3.58 \times 10^{-2}}{\hbar^{2} \eta}} \leq \hbar \sqrt{\frac{3.58 \times 10^{-2}}{\hbar^{2}}} \approx 1.9 \times 10^{-4} \textit{m}.
\end{equation}
The substitution of the root (\ref{r2}) hold the inequality (\ref{EKMMGUP}) for all $\beta > 0$.

Consequently, the upper bound for the minimal-length value is order $10^{-4}$ m. Therefore, using the experiment described above a result with less restrictions than those reported in the literature is obtained \cite{Nogueira:2016,Vagenas1:2022,Sen:2022,Saha1:2020,Gusson:2018,Pikovski2:2012,Nogueira:2013}.


\section{Conclusion}
\label{Concl}

In this work we find the non-zero minimal uncertainties induced by the main proposals of GUP's (KMM, ADV, Pedram and Nouicer) which are modified due to the quantum entanglement of a system of two identical particles. In principle, our results have pointed out that the minimal uncertainties are reduced by half for a system of two entangled identical particles independently of the GUP. Hence, if a non-zero minimal uncertainty in position can be interpreted as a minimal length then the quantum entanglement reduces by half the minimal length. However, the minimal length must not depend on the physical system. We overcome  this apparent paradox  by using the Quesne and Tkachuck's proposal for a system composed. Consequently, despite the quantum entanglement to change the GUP, the minimal length does not change.  Based on our results and using the reference \cite{Rigolin:2016} we can expect that the apparent minimal uncertainty for a system of $N$ entangled identical particles is reduced by $\frac{1}{N}$, nonetheless the minimal length does not change because the effective parameter $\beta$ is  also reduced by a factor $\frac{1}{N^{2}}$.

Finally, we have estimated from the data obtained from the Kim and Shih's experiment an upper bound value for the minimal length of the order of $10^{-4}$ m. Consequently, it is rather an inexpressive value (in the sense of leading to poor predictive power) as compared to ones have been found in the literature. This is due to the high imprecision of the experiment on the entangled system described above. However, we may expect that more refined version of the experiment may lead to more stringent bounds on the minimum length.


\section*{Acknowledgments}

We would like to thank FAPES, CAPES and CNPq (Brazil) for financial support.

\end{document}